\documentclass[letterpaper, 10 pt, conference]{ieeeconf}

\IEEEoverridecommandlockouts  
                                                         
\usepackage{graphicx}
\usepackage{amssymb,amsmath,dsfont,amsfonts}
\usepackage{epstopdf}
\usepackage{booktabs}
\usepackage{tikz}
\usepackage{cite}
\usepackage{color}
\usepackage{boxedminipage}
\usepackage{subfigure}
\usepackage{algorithmic}
\usepackage{algorithm}
\usepackage{multirow,bigdelim}
\usepackage{multirow}
\usepackage{url}

\newtheorem{definition}{Definition}

\newtheorem{remark}{Remark}

\newcounter{l1}
\newcommand{\barablist}{\begin{list}{\arabic{l1}}{\usecounter{l1}}}

\usepackage{acronym}

\title{\LARGE \bf
Low-dimensional Models in Spatio-Temporal Wind Speed Forecasting
}

\author{Borhan M. Sanandaj$\text{i,}^{\star}$ Akin Tascikaraogl$\text{u,}^{\diamond,\star}$, Kameshwar Pooll$\text{a,}^{\star}$ and Pravin Varaiy$\text{a}^{\star}$
\thanks{\textsuperscript{$\star$}B. M. Sanandaji, A. Tascikaraoglu, K. Poolla, and P. Varaiya are all with Department of Electrical Engineering and Computer Sciences, University of California, Berkeley, CA 94720, USA. 
Email: \{sanandaji,atasci,poolla,varaiya\}@berkeley.edu.}
\thanks{\textsuperscript{$\diamond$} A. Tascikaraoglu is also with the Department of Electrical Engineering, Yildiz Technical University, Istanbul, Turkey. 
Email: atasci@yildiz.edu.tr.}
\thanks{Supported in part by EPRI and CERTS under sub-award 09-206; PSERC S-52; NSF under Grants 1135872, EECS-1129061, CPS-1239178, CNS-1239274; the Republic of Singapore National Research Foundation through a grant to the Berkeley Education Alliance for Research in Singapore for the SinBerBEST Program; Robert Bosch LLC through its Bosch Energy Research Network funding program;
TUBITAK-2219 Program.}
}

\def\real{ \mathbb{R} }

\newcommand{\vc}[1]{\boldsymbol{#1}}

\acrodef{CoM}{Concentration of Measure}
\acrodef{i.i.d.}{independent and identically distributed}
\acrodef{LTI}{Linear Time-Invariant}
\acrodef{LTV}{Linear Time-Variant}
\acrodef{LPV}{Linear Parameter-Varying}
\acrodef{RIP}{Restricted Isometry Property}
\acrodef{SVD}{Singular Value Decomposition}
\acrodef{CS}{Compressive Sensing}
\acrodef{DSP}{Digital Signal Processing}
\acrodef{CSI}{Compressive System Identification}
\acrodef{CTI}{Compressive Topology Identification}
\acrodef{CBD}{Compressive Binary Detection}
\acrodef{OMP}{Orthogonal Matching Pursuit}
\acrodef{MP}{Matching Pursuit}
\acrodef{ERC}{Exact Recovery Condition}
\acrodef{BOMP}{Block Orthogonal Matching Pursuit}
\acrodef{COMP}{Clustered Orthogonal Matching Pursuit}
\acrodef{CoSaMP}{Compressive Sampling Matching Pursuit}
\acrodef{KKT}{Karush-Kuhn-Tucker}

\acrodef{FIR}{Finite Impulse Response}
\acrodef{DFT}{Discrete Fourier Transform}
\acrodef{DCT}{Discrete Cosine Transform}
\acrodef{JL}{Johnson-Lindenstrauss}
\acrodef{ROC}{Receiver Operating Curve}
\acrodef{NP}{Neyman-Pearson}
\acrodef{ARX}{Auto Regressive with eXternal input} 
\acrodef{MISO}{Multi-Input Single-Output}
\acrodef{SISO}{Single-Input Single-Output}
\acrodef{MIMO}{Multi-Input Multi-Output}
\acrodef{BP}{Basis Pursuit}
\acrodef{LASSO}{Least Absolute Shrinkage and Selection Operator}
\acrodef{GLASSO}{Group LASSO}
\acrodef{NNG}{Non-Negative Garrote}
\acrodef{LARS}{Least Angle Regression}
\acrodef{I/O}{Input/Output}
\acrodef{CST-WSF}{Compressive Spatio-Temporal Wind Speed Forecasting}
\acrodef{AR}{Autoregressive}
\acrodef{M-AR}{Multivariate Autoregressive}
\acrodef{NM-AR}{Nonuniform Multivariate Autoregressive}
\acrodef{RPS}{Renewable Portfolio Standard}
\acrodef{NWP}{Numerical Weather Prediction}
\acrodef{TDD}{Trigonometric Direction Diurnal}
\acrodef{RSTD}{Regime Switching Space-Time Diurnal}
\acrodef{TDDGW}{TDD with Geostrophic Wind Information}
\acrodef{ACK}{Nantucket Memorial Airport}
\acrodef{WT}{Wavelet Transform}
\acrodef{ANN}{Artificial Neural Network}
\acrodef{MAE}{Mean Absolute Error}
\acrodef{RMSE}{Root Mean Squared Error}
\acrodef{NRMSE}{Normalized Root Mean Squared Error}
\acrodef{TSOs}{Transmission System Operators}
\acrodef{IPPs}{Independent Power Producers}
\acrodef{ST-ANN}{Spatio-Temporal ANN}
\acrodef{LS}{Least Squares}
\acrodef{METAR}{Meteorological Terminal Aviation Routine}

\usepackage{color}

\newcommand{\cut}[1]{}

\begin{document}

\maketitle
\thispagestyle{empty}
\pagestyle{empty}


\begin{abstract}

Integrating wind power into the grid is challenging because of its random nature.  Integration is facilitated with accurate short-term forecasts of wind power. The paper presents a \emph{spatio-temporal} wind speed forecasting algorithm that incorporates the time series data of a target station and data of surrounding stations.
Inspired by Compressive Sensing (CS) and structured-sparse recovery algorithms, we claim that there usually exists an intrinsic low-dimensional structure governing a large collection of stations that should be exploited.
We cast the forecasting problem as recovery of a \emph{block-sparse} signal $\vc{x}$ from a set of linear equations 
$\vc{b} = A\vc{x}$ for which we propose novel structure-sparse recovery algorithms.
Results of a case study in the east coast show that the proposed Compressive Spatio-Temporal Wind Speed Forecasting (CST-WSF) algorithm significantly improves the short-term forecasts compared to a set of widely-used benchmark models.
\end{abstract}

\acresetall
\section{Introduction}

\subsection{Variable Energy Resources}
Many countries in the world as well as many states in the U.S. have mandated aggressive \acp{RPS}. Among different renewable energy resources, wind energy itself is expected to grow to provide between $15$ to $25\%$ of the world's global electricity by 2050. According to another study, the world total wind power capacity has doubled every three years since 2000, reaching an installed capacity of 197 GW in 2010 and 369 GW in 2014~\cite{iea2013technology,CA_renewable_portfolio}. 
The random nature of wind, however, makes it difficult to achieve the power balance needed for its integration into the grid~\cite{smith2007utility,bitar2011role}. 
The use of ancillary services such as frequency regulation and load following  to compensate for such imbalances\cite{callaway2009tapping,tascikaraoglu2011power,sanandaji2014fast,qin2014modeling} is facilitated by accurate forecasts~\cite{wang2008security, tascikaraoglu2014adaptive}.


\subsection{Wind Energy Forecasting Methods} 

One can directly attempt to forecast wind power.~An alternative approach is to  forecast the wind speed and then convert it to wind power using given power curves. This approach will accommodate different wind turbines installed in a wind farm  experiencing the same wind speed but resulting in different wind power generation. We focus on wind speed forecasting in this paper. 
Wind speed forecasting methods can be categorized to different groups: (i) model-based methods such as \ac{NWP} vs. data-driven methods, (ii) point forecasting vs. probabilistic forecasting, and (iii) short-term forecasting vs. long-term forecasting.  This paper is concerned with short-term point forecasting using both temporal data as well as spatial information. For a more complete survey of wind speed forecasting methods see~\cite{zhu2012short} and~\cite{tascikaraoglu2014review} among others.

\begin{figure*}
\begin{equation}
\underbrace{
\left[
\begin{array}{c}
y_{n+1}^i\\
y_{n+2}^i\\
\vdots\\
y_{n+M}^i
\end{array}
\right]}_{\vc{b}\in \real^{M}}=
\underbrace{
\left[
\begin{array}{ccc|cc|ccc}
y_{n}^1 & \hdots & y_{1}^1 & \hdots & \hdots &y_{n}^P & \hdots & y_{1}^P\\
y_{n+1}^1  & \ddots & \vdots & & & y_{n+1}^P & \ddots & \vdots \\
\vdots & \ddots & \vdots & & & \vdots & \dots & \vdots\\
y_{n+M-1}^1 &  \hdots & y_{M}^1 & \hdots & \hdots & y_{n+M-1}^P  & \hdots & y_{M}^P \\
\end{array}
\right]}_{A \in \real^{M \times N}}
\underbrace{
\left[
\begin{array}{c}
{X}_{1}^{\text{tr}}(1,i)\\
\vdots\\
{X}_{n}^{\text{tr}}(1,i)\\[1ex]
\hline \\[-1.5ex]
\vdots\\
\vdots\\[1ex]
\hline \\[-1.5ex]
{X}_{1}^{\text{tr}}(P,i)\\
\vdots\\
{X}_{n}^{\text{tr}}(P,i)\\
\end{array}
\right]}_{{\vc{x} \in \real^{N}}}
\begin{array}{l}\\[-12mm] \rdelim\}{4}{6mm}[$\text{Block}~1$] \\ \\ \\ \\ \\ \\[7mm] \rdelim\}{4}{6mm}[$\text{Block}~P$] \\ \\
\end{array} \\[-1ex]
\label{eq:X_vector}
\end{equation}
\end{figure*}

\subsection{Spatio-Temporal Wind Speed Forecasting}

There is a growing interest in the so-called \emph{spatio-temporal} forecasting methods that use information from neighboring stations to improve the forecasts  of a target station, since there is a significant cross-correlation between the time series data of a target station and its surrounding stations. We review some of the spatio-temporal forecasting methods.
Gneiting et al.~\cite{gneiting2006calibrated} introduced the \ac{RSTD} model for average wind speed data based on both spatial and temporal information. This method was later improved by Hering and Genton~\cite{hering2010powering} who incorporated  wind direction in the forecasting process by introducing \ac{TDD} model. 
Xie et al.~\cite{xieshort} also considered probabilistic \ac{TDD} forecast for power system economic dispatch. 
Dowell et al.~\cite{dowell2013short} employed a multi-channel adaptive filter to predict the wind speed and direction by taking advantages of spatial correlations at numerous geographical sites.
He et al.~\cite{he2014spatio} presented Markov chain-based stochastic models for predictions of wind power generation after characterizing the statistical distribution of aggregate power with a graph learning-based spatio-temporal analysis.
Regime-switching models based on wind direction are studied by 
Tastu et al.~\cite{tastu2011spatio} where they consider various statistical models, such as ARX models, to understand the effects of different variables on forecast error  characteristics.
A methodology with probabilistic wind power forecasts in the form of predictive densities taking the spatial information into account was developed in~\cite{tastu2014probabilistic}. 
Sparse Gaussian Conditional Random Fields (CRFs) have also been deployed for probabilistic wind power forecasting~\cite{wytock2013largescale}.
See~\cite{zhang2014review} for a comprehensive review of the state-of-the-art methods.

\subsection{Our Contribution}

Inspired by Compressive Sensing (CS) and structured-sparse recovery algorithms, we claim that there usually exists an intrinsic low-dimensional structure governing the interactions among a large collection of weather stations. Such low-dimensional models should be exploited in the forecasting process.
To this end, we cast the forecasting problem as the recovery of a \emph{block-sparse} signal $\vc{x}$ from a set of linear equations $\vc{b} = A\vc{x}$ for which we propose novel structure-sparse recovery algorithms.
As a case study, we apply our proposed forecasting algorithm to the data recorded from 57 measuring locations (a combination of airports and weather stations) in a region in the east coast including Massachusetts, Connecticut, New York, and New Hampshire. The results lead to a considerable improvement of the short-term forecasts compared to a set of widely-used benchmark models and advanced spatio-temporal approaches.


\subsection{Paper Organization}
In Section~\ref{sec:MAR}, we formulate the forecasting problem. The proposed forecasting algorithms and  related concepts are presented in Section~\ref{sec:CST-WSF}. We apply the proposed methods to real wind speed data and compare the results with other benchmark methods in Section~\ref{sec:case}. Section~\ref{sec:conc} presents our concluding remarks and possible future directions.


\section{Multivariate Autoregressive (M-AR) Model}
\label{sec:MAR}

\ac{AR} models assume that the output variable of a system can be well presented as a weighted linear combination of its own previous values.~\ac{M-AR} (a.k.a., Vector Autoregressive) models generalize this approach to multivariate time series.~Let $\vc{y}(t) \in \real^P$ be a $P$-dimensional output measurement (e.g., wind speeds at $P$ weather stations) at time $t$. An \ac{M-AR} model of order $n$ is written as 
\begin{equation}
\begin{split}
\vc{y}(t)&=X_1\vc{y}(t-1)+\cdots+X_n\vc{y}(t-n)+\vc{e}(t)\\
&=\sum_{j=1}^n{X_j\vc{y}(t-j)+\vc{e}(t)}, 
\end{split}
\label{eq:M_AR_model}
\end{equation}
where $X_j \in \real^{P \times P}$ is a coefficient matrix associated with the $j$-th time lag and $\vc{e}(t)$ is a Gaussian noise.
Using a different notation, let $y^i_t$ be the wind speed of the $i$-th station at sample time $t$ for $t = 1,2,\dots,M+n$. For each station, the \ac{M-AR} model~(\ref{eq:M_AR_model}) can be re-written in a matrix-vector product format as  in
(\ref{eq:X_vector}),
where $N := nP$.
In the training stage, the goal is to find a coefficient vector $\vc{x} \in \real^N$ that best explains  the observations $\vc{b} \in \real^M$ and $A \in \real^{M \times N}$.
As  seen from~(\ref{eq:X_vector}), $\vc{x}$ has a block structure as the coefficients corresponding to each station appear in one vector-block.

\section{Compressive Spatio-Temporal \\Wind Speed Forecasting (CST-WSF)}
\label{sec:CST-WSF}
%
We believe that among a large collection of stations, only a few of them have a strong correlation with the target station. We show that under the assumption of {\em sparsity} of the  interconnections (that is, assuming only a few weather stations contribute to the output of the target station), there will be a distinct structure to the solution $\vc{x}$ that we are seeking. In particular, a typical coefficient vector $\vc{x}$ under our model assumptions will have very few non-zero entries, and these non-zero entries will be clustered in few locations. Vectors with such structure are called~\emph{block-sparse}. The number of blocks corresponds to the number of links that contribute to the output of the target station. For a given target station, we then solve the  minimization problem:
\begin{equation}
\min_{\vc{x}} \|\vc{b}-A\vc{x}\|_2 \ \ \ \text{subject to} \ \ \ (\vc{x} \ \text{is block-sparse}).
\label{eq:find_eq}
\end{equation}
We call this approach~\ac{CST-WSF}, as it is inspired by \ac{CS} and structured-sparse recovery.
\subsection{Background on \ac{CS}}

\ac{CS} enables recovery of an unknown signal from its underdetermined set of measurements under the assumption of sparsity of the signal and under certain conditions on the measurement matrix $A$~\cite{candes2006robust}. 
The \ac{CS} recovery problem can be viewed as recovery of a $K$-sparse signal $\vc{x} \in \real^N$ from its observations $\vc b = A\vc x \in \real^M$ where $A \in \real^{M \times N}$ is the measurement matrix with $M < N$ (in many cases $M \ll N$). 
A $K$-sparse signal $\vc{x} \in \real^N$ is a signal of length $N$ with $K$ non-zero entries where $K < N$. Since the null space of $A$ is non-trivial, there are infinitely many candidate solutions to the equation $\vc{b} = A\vc{x}$; however, \ac{CS} recovery algorithms exploit the fact that, under certain conditions on $A$, only one candidate solution is suitably sparse. 
An interested reader can refer to~\cite{tropp2004greed,donoho2001uncertainty} for several proposed recovery conditions.

%
\subsection{Uniform \ac{CST-WSF}}
We adapt our \ac{CST-WSF} algorithm from tools proposed in the \ac{CS} literature for recovery of a block-sparse signal $\vc{x}$.
\begin{definition} [Block $K$-Sparse Signal] 
Let $\vc{x} \in \real^N$ be a concatenation of $P$ vector-blocks $\vc{x}_i \in \real^{n}$, i.e.,
\begin{equation*}
\vc{x}=[\vc{x}^\text{tr}_1 \cdots \vc{x}^\text{tr}_i \cdots \vc{x}^\text{tr}_P]^\text{tr},
\end{equation*}
where $N = nP$. A signal $\vc x \in \real^N$ is called \emph{block $K$-sparse} if it has $K<P$ non-zero blocks. \hfill $\square$
\label{def:block}
\end{definition} 

Several extensions of the standard \ac{CS} recovery algorithms can account for additional structure in the sparse signal to be recovered~\cite{baraniuk2010mbcs,eldar2010block}. 
Among these, the \ac{BOMP} algorithm~\cite{eldar2010block} is designed to exploit block sparsity due to its flexibility in recovering block-sparse signals of different sparsity levels and its low computation complexity~\cite{sanandaji2012review}. In a more general setting, \ac{BOMP} has been recently used for topology identification of interconnected dynamical systems~(e.g., see \cite{sanandaji2012thesis}).

\begin{figure*}
\begin{equation}
\small
\underbrace{
\left[
\begin{array}{c}
y_{n_{\text{max}}+1}^i\\
y_{n_{\text{max}}+2}^i\\
\vdots\\
y_{n_{\text{max}}+M}^i
\end{array}
\right]}_{\vc{b}\in \real^M}=
\underbrace{
\left[
\begin{array}{ccccc|c|ccc}
y_{n_{\text{max}}}^1 & \hdots & \hdots & \hdots & y_{n_{\text{max}}-n_1+1}^1 & \hdots &y_{n_{\text{max}}}^P& \hdots & y_{n_{\text{max}}-n_P+1}^P\\
y_{n_{\text{max}}+1}^1  & \ddots & \ddots & \ddots & \vdots & &y_{n_{\text{max}}+1}^P & \ddots & \vdots \\
\vdots & \ddots & \ddots & \ddots & \vdots & & \vdots & \ddots & \vdots\\
y_{n_{\text{max}}+M-1}^1 &  \hdots  & \hdots & \hdots & y_{n_{\text{max}}-n_1+M}^1 & \hdots & y_{n_{\text{max}}+M-1}^P & \hdots &y_{n_{\text{max}}-n_P+M}^P \\
\end{array}
\right]}_{A \in \real^{M \times N}}
\underbrace{
\left[
\begin{array}{c}
{X}_{1}^{\text{tr}}(1,i)\\
\vdots\\
\vdots\\
\vdots\\
{X}_{n_1}^{\text{tr}}(1,i)\\[1ex]
\hline \\[-1.5ex]
\vdots\\[1ex]
\hline \\[-1.5ex]
{X}_{1}^{\text{tr}}(P,i)\\
\vdots\\
{X}_{n_P}^{\text{tr}}(P,i)\\
\end{array}
\right]}_{{\vc{x} \in \real^N}}
\begin{array}{l}\\[-13mm] \rdelim\}{7}{6mm}[$\text{Block}~1$] \\ \\ \\ \\ \\[19mm] \rdelim\}{4}{6mm}[$\text{Block}~P$] \\ \\
\end{array} \\[-1ex]
\label{eq:X_vector_nonuniform}
\end{equation}
\end{figure*}
%


\subsection{Nonuniform \ac{CST-WSF}}

In a uniform \ac{CST-WSF}, the assumption is that a uniform \ac{M-AR} model as given in (\ref{eq:X_vector}) is governing the interactions between stations. In other words, we assume that  the target station and its surrounding stations are related by \ac{AR} models of the same order. In this section, we consider a more generalized version of the \ac{CST-WSF} algorithm where the target station and its surrounding stations are related by \ac{AR} models of different orders. This model structure, called \ac{NM-AR} distinguishes between the stations with high and low cross-correlation with the target station. Let $n_i$ be the order associated with the $i$-th station for $i=1,2,\dots,P$. An \ac{NM-AR} version of (\ref{eq:X_vector}) can be considered as given in (\ref{eq:X_vector_nonuniform}), where $n_{\text{max}} \geq \max_i n_i$ and $N := \sum_{i=1}^Pn_i$. This  results in a \emph{nonuniform} block-sparse coefficient vector $\vc{x}$ whose blocks have different length.  
\begin{definition} [Nonuniform Block $K$-Sparse Signal] Let $\vc{x} \in \real^N$ as a concatenation of $P$ vector-blocks $\vc{x}_i \in \real^{n_i}$
%
%
where $N = \sum_{i=1}^Pn_i$. A signal $\vc x \in \real^N$ is called \emph{nonuniform block $K$-sparse} if it has $K<P$ non-zero blocks. \hfill $\square$
\label{def:block_nonuni}
\end{definition} 
\begin{remark}
One should note that our definition of nonuniform block $K$-sparse signals is a generalization of the conventional definition of block $K$-sparse signals (Definition \ref{def:block}) where all blocks have the same length, i.e., 
$n_i=n, \forall i$. \hfill $\square$
\end{remark}

Given $\{n_i\}_{i=1}^{P}$, the \ac{BOMP} algorithm can be used for recovery of $\vc{x}$ with $A_i\in \real^{M \times n_i}$. In order to find the set
of order $\{n_i\}_{i=1}^{P}$, we use a correlation analysis. We  then adjust the orders to achieve the best prediction performance.



\section{Case Study of 57 Stations in East Coast}
\label{sec:case}
We apply our proposed \ac{CST-WSF} algorithms to real wind speed data. East coast states are good candidates for our study because: (i) wind speed profiles are higher and (ii) there are more stations in a close vicinity in these states.  

\subsection{Data Description}

\begin{figure}[tb]
\begin{center}
 \includegraphics[width =1\columnwidth]{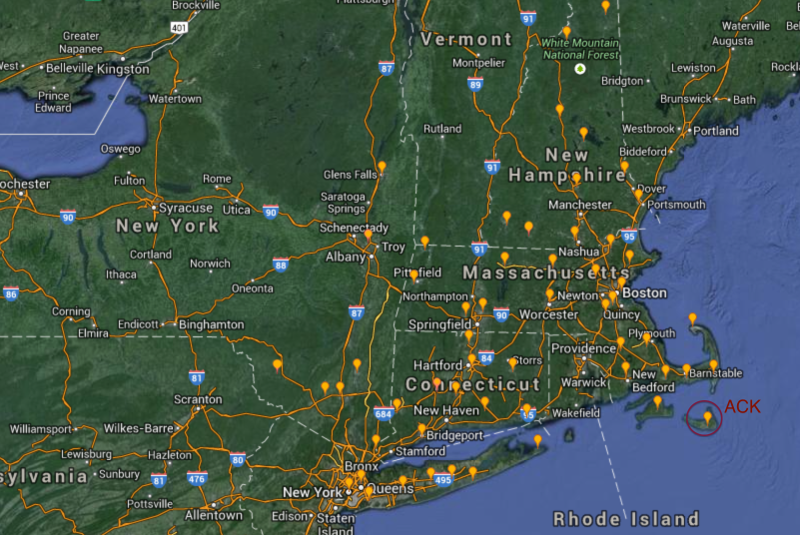}
\end{center}
\caption{Map of the area under study. The $57$ measuring locations in east coast are shown with yellow points. Circled in red is the target station ACK.}
\label{fig:stations_map}
\end{figure}

We use hourly wind speed data from \ac{METAR} weather reports of 57 stations in east coast including Massachusetts, Connecticut, New York, and New Hampshire~\cite{Iowa}. 
Fig.~\ref{fig:stations_map} depicts the area under study and the location of these 57 stations. 
The target station~\ac{ACK} (circled in red) is located on an island and is subject to wind profiles with high ramps and speeds due to the fact that the surrounding surface has very low roughness heights. Furthermore, this area has good correlations with other stations owing to the fact the prevailing wind direction of this region is mainly northwest or southeast. 
A time period from January 6, 2014 to February 20, 2014 is considered in our simulations. This time period has the most unsteady wind conditions throughout the year. The data is divided to two subsets: (i) training subset from January 6, 2014 to February 6, 2014 (a period of 30 days) and (ii) validation subset from February 6, 2014 to February 20, 2014 (a period of 14 days).

\subsection{Comparison with Other Benchmark Algorithms}

\begin{figure}[ht!]
\centering
\subfigure[Persistence forecasting]{
   \includegraphics[width = 0.825\columnwidth] {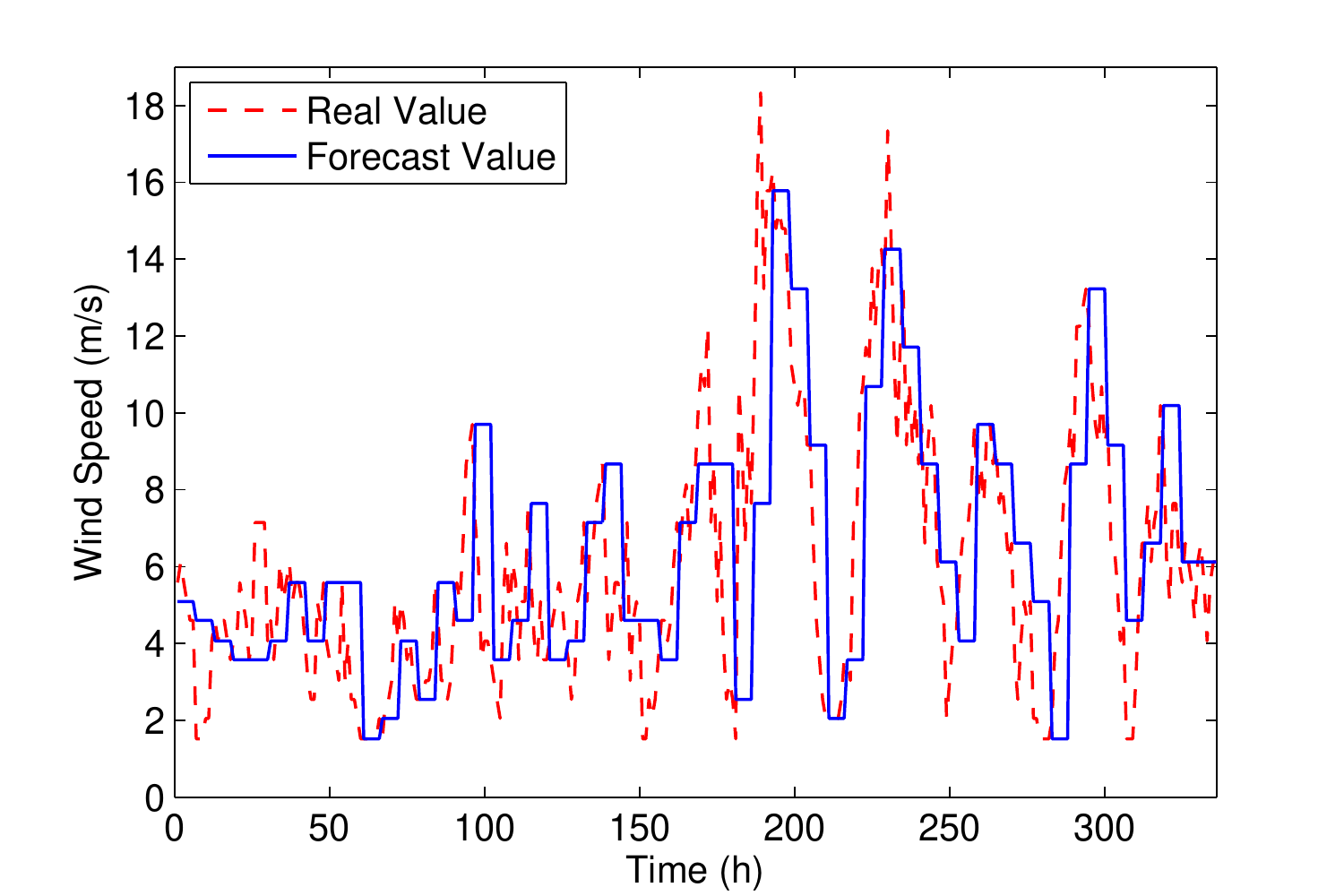}
   \label{fig:persistence}
 }
 \subfigure[AR model of order 3]{
   \includegraphics[width = 0.825\columnwidth] {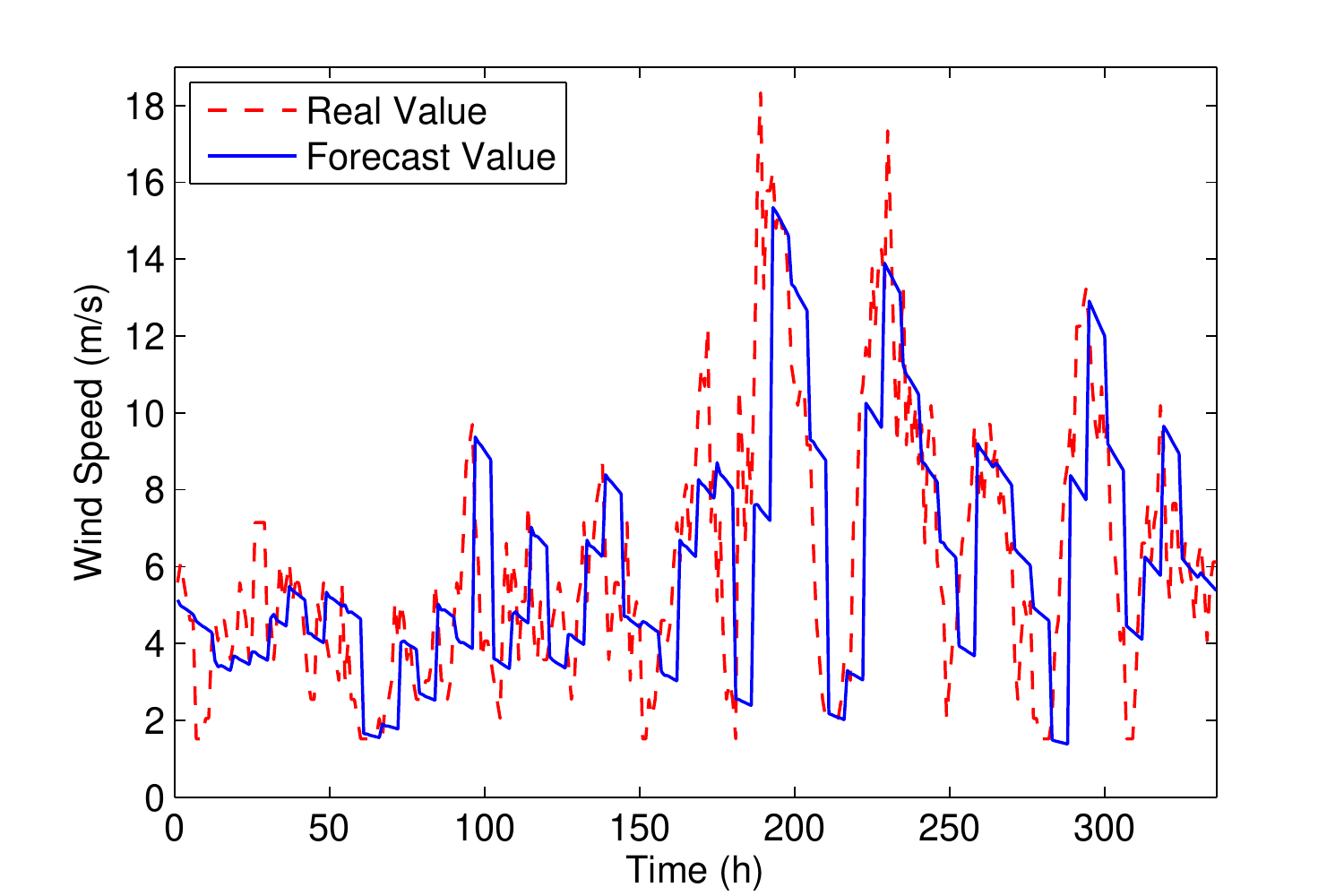}
   \label{fig:AR3}
 }
 \subfigure[WT-ANN model]{
   \includegraphics[width = 0.825\columnwidth] {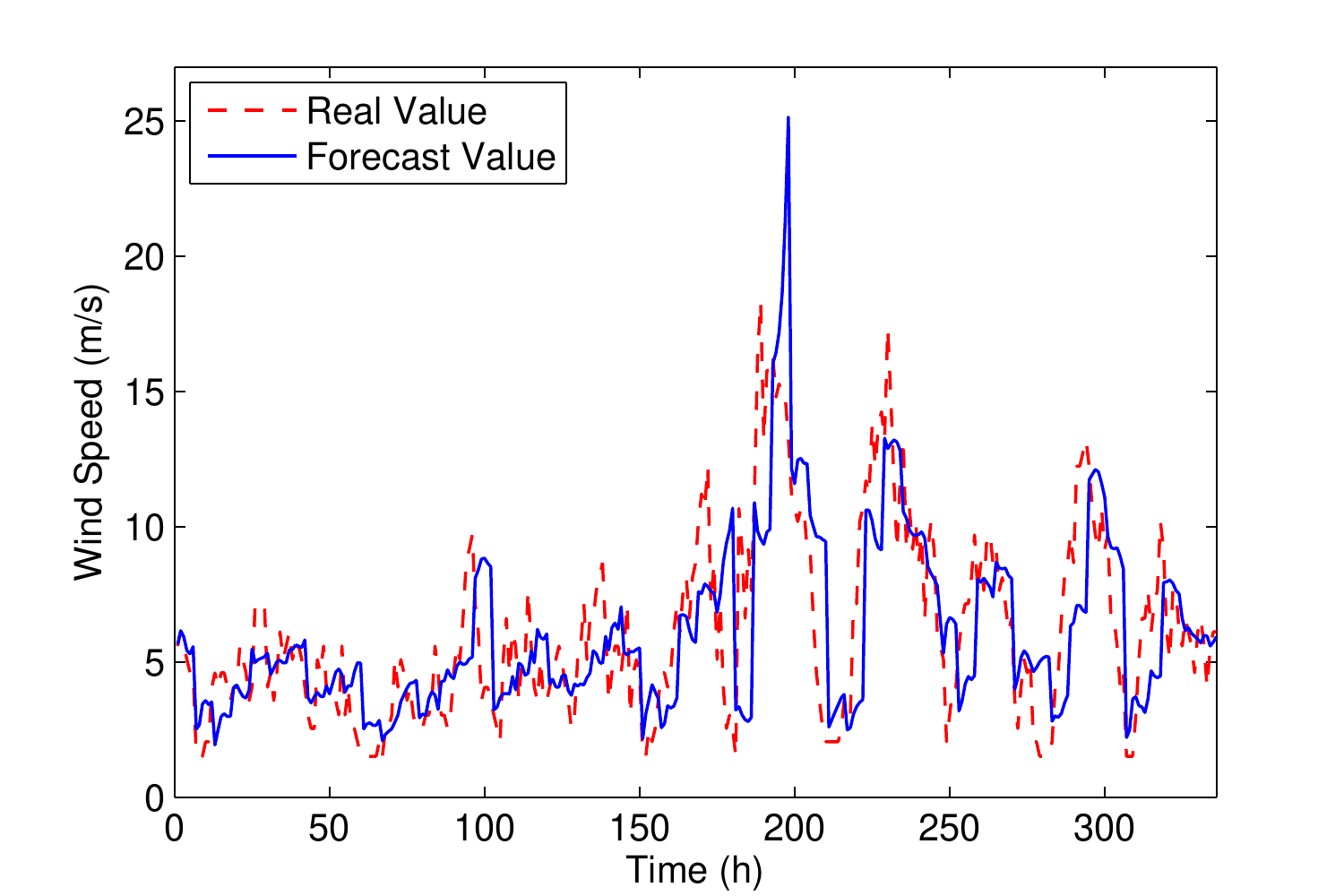}
   \label{fig:WT_ANN}
 }
\caption{Comparison of different temporal forecasting algorithms.}
\label{fig:comparisons}
\end{figure}

In order to better gauge the effectiveness of the proposed algorithm, we compare \ac{CST-WSF} with other proposed benchmark algorithms in temporal and 
spatio-temporal wind speed forecasting.
For temporal wind speed forecasting, we first consider~\emph{persistence forecasting} method which simply uses the last measured value for the forecast interval. Any algorithm that can improve upon persistence forecasting is judged to be an effective algorithm. 
We also consider \ac{AR} models as well as an advanced prediction model that combines \ac{WT} with \ac{ANN}. The latter method is shown to have the capability of capturing nonlinearity in the wind speed time series. In this model, briefly, the volatile wind speed series is first cut up by the \ac{WT}  into a number of better-behaved sub-series in various frequency bands. Subsequently, estimates of each extracted sub-series are carried out separately employing the \ac{ANN} mode. Speed predictions are then reconstructed except for the highest frequency band which represents the most fluctuating part of the wind speed series (see~\cite{tascikaraoglu2012assessment} for more details). 
Multi-step predictions were performed in a recursive manner for a period of  14 consecutive days with 6 hour-ahead updates.
The prediction results are given in Fig. \ref{fig:comparisons}.
As can be seen, the considered temporal prediction methods provide reasonable forecasting compared to persistence model. \ac{ANN}-based model outperforms the \ac{AR} model. 

\begin{figure}[ht!]
\centering
\subfigure[Spatio-temporal ANN model]{
  \includegraphics[width = 0.825\columnwidth]{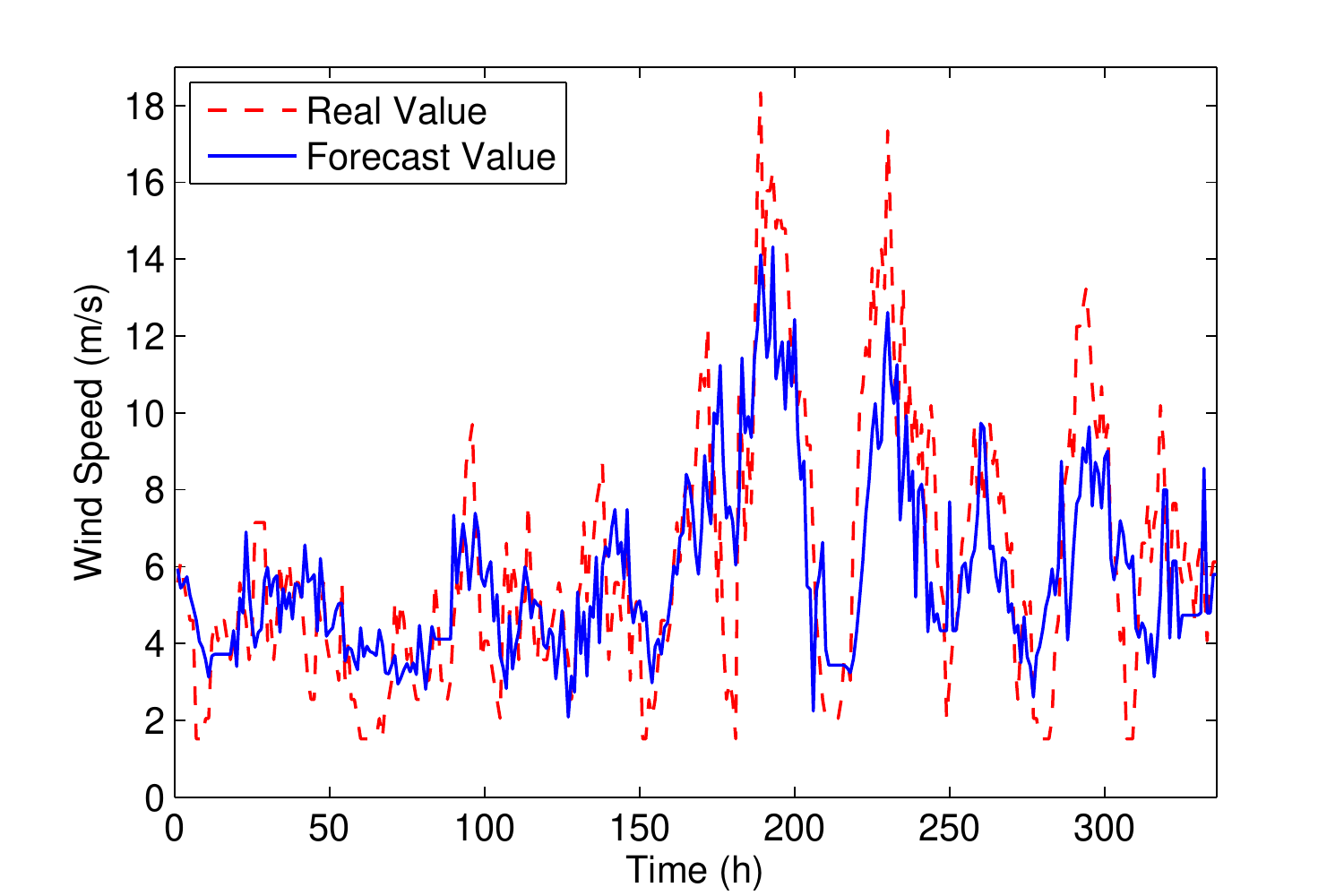}
   \label{fig:ST_ANN}
 }
 \subfigure[Spatio-temporal LS M-AR model]{
  \includegraphics[width = 0.825\columnwidth] {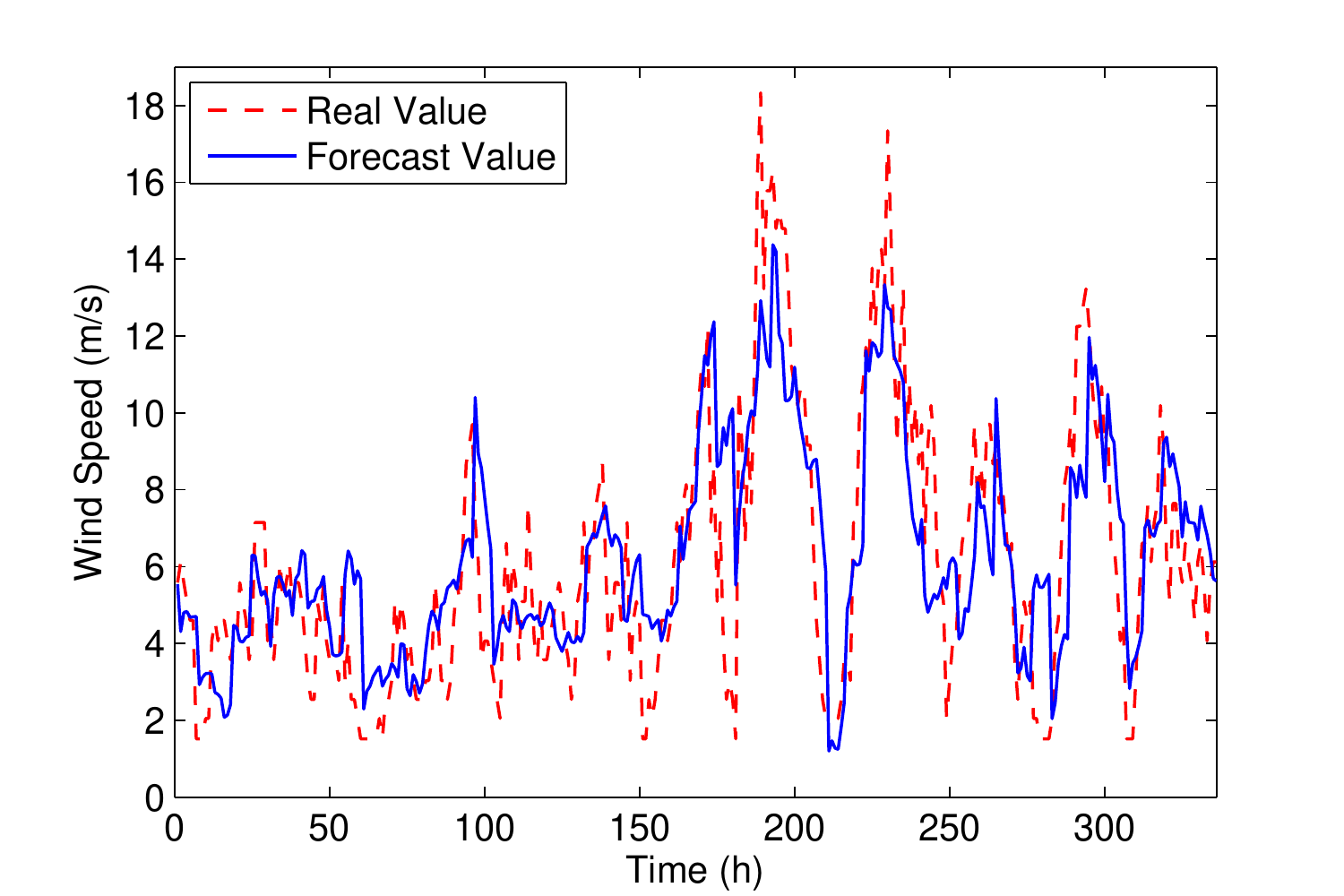}
   \label{fig:LS}
 }
\caption{Comparison of different spatio-temporal forecasting algorithms.}
\label{fig:spatial_comparisons}
\end{figure}

We also consider two spatio-temporal forecasting methods. We first employed an advanced \ac{ANN}-based spatio-temporal model~\cite{bilgili2007application}. We also employed a \ac{LS} \ac{M-AR} spatio-temporal forecasting approach~\cite{tastu2011spatio}. Fig.~\ref{fig:spatial_comparisons} depicts how incorporation of spatial information improves the forecasting performance as compared to temporal methods. 

\begin{figure}[ht!]
\centering
\subfigure[Uniform CST-WSF with equal orders of 3.]{
   \includegraphics[width = 0.825\columnwidth] {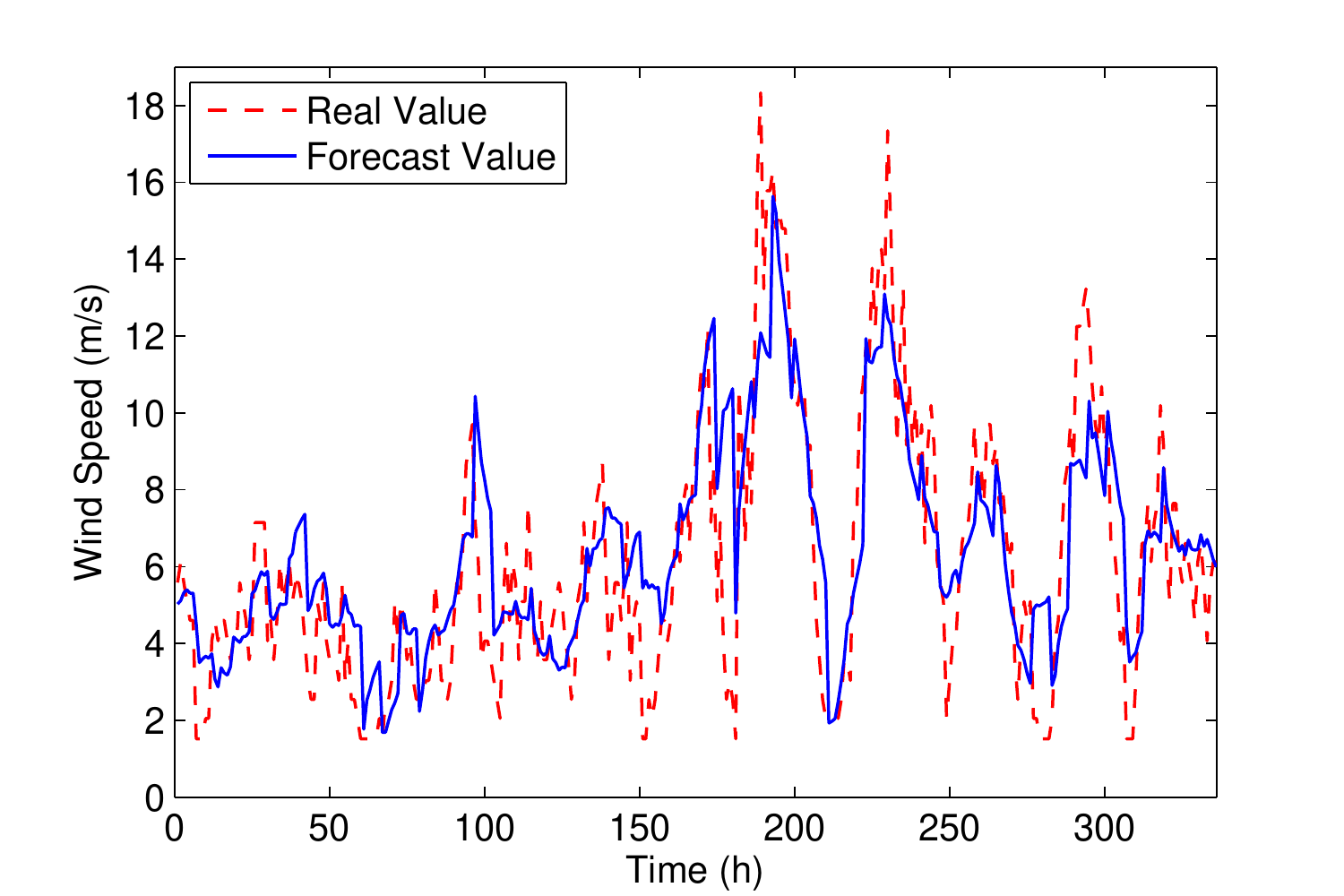}
   \label{fig:BOMP}
 }
 \subfigure[Nonuniform CST-WSF with nonuniform orders.]{
   \includegraphics[width = 0.825\columnwidth] {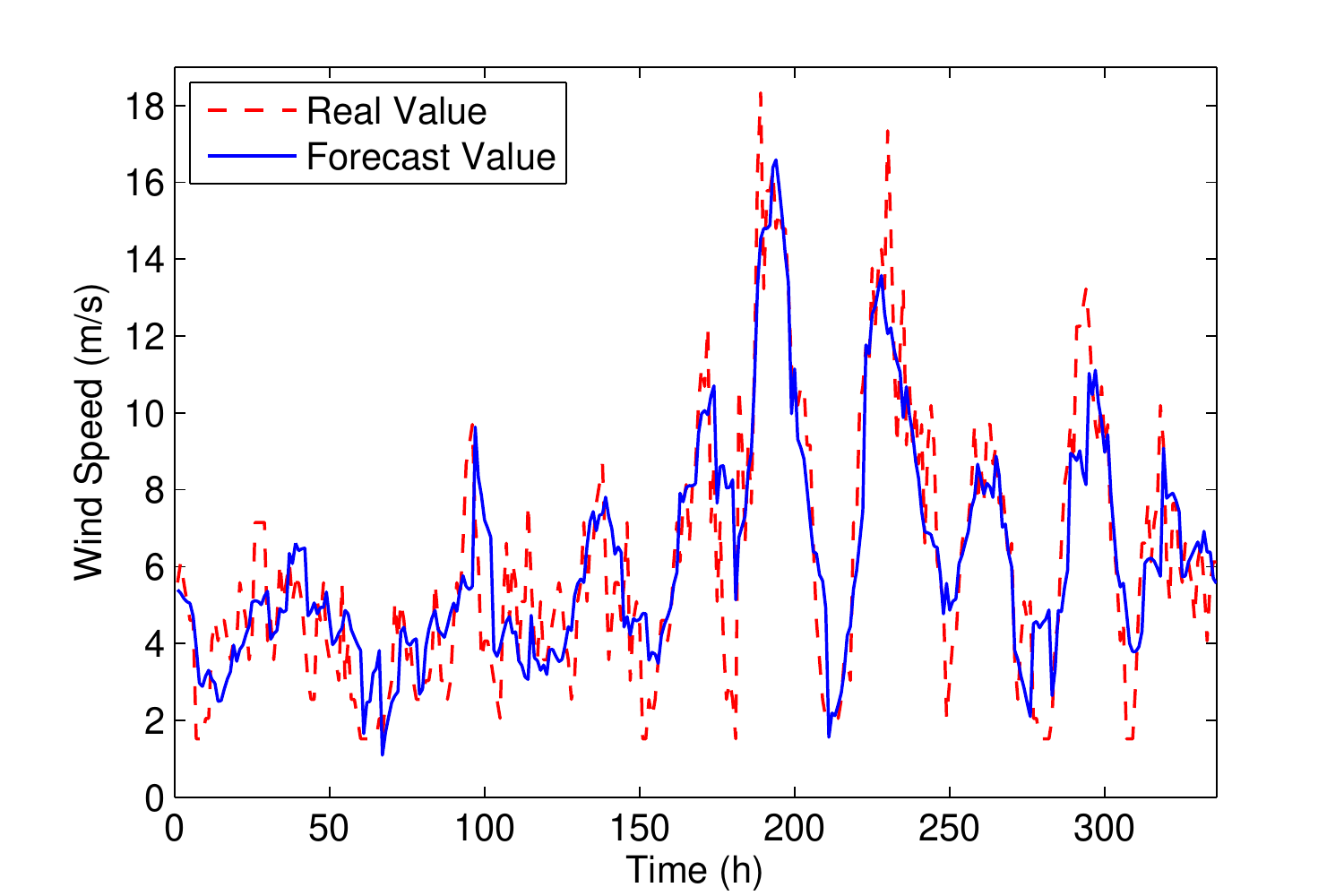}
   \label{fig:nonuniform_BOMP}
 }
\caption{Comparison of the uniform and nonuniform CST-WSF.}
\label{fig:BOMP_models}
\end{figure}

\subsection{Uniform and Nonuniform \ac{CST-WSF}}

We now apply our proposed uniform and nonuniform \ac{CST-WSF} methods. Note that in our simulations a new coefficient vector $\vc{x}$ is obtained every 6 hours (equivalently, every 6 time steps as each time step is 1 hour). This is the considered prediction time horizon. 
Also, we follow a recursive approach in prediction. That is, the wind speed predictions at time $n+M+1$ for all stations ($\hat{y}^i_{n+M+1}, \forall i$) are included in the $A$ matrix for predicting the wind speed at time $n+M+2$ ($\hat{y}^i_{n+M+2}, \forall i$) and so on. This recursive process goes on for 6 time steps.
The elements of $A$ are then completely updated with real measurements and the recursive process continues for another 6 time steps.

Figure~\ref{fig:BOMP} shows the result of the uniform 
\ac{CST-WSF}. The result is superior to all benchmark approaches discussed in the previous section. 
We then apply the nonuniform \ac{CST-WSF} algorithm. The result is illustrated in Fig.~\ref{fig:nonuniform_BOMP}. 
In order to demonstrate the effectiveness of the proposed forecasting models, the associated \ac{MAE} and \ac{RMSE} values, which are the most common performance metrics in the wind forecasting literature, are listed in Table \ref{error measures} for all of the wind speed forecasting methods considered in this paper. The \ac{MAE} provides the average deviation between the measured and predicted data while the \ac{RMSE} gives higher weights to larger error values by squaring the differences.  
Moreover, the \ac{NRMSE} which is the \ac{RMSE} normalized by the range of observed data, is calculated to provide a scale-independent error measure.
Evidently the proposed uniform and nonuniform \ac{CST-WSF} methods outperform
the other considered temporal and spatio-temporal methods. The best prediction is provided by the nonuniform \ac{CST-WSF}. Considering the \ac{NRMSE} as an example,  nonuniform \ac{CST-WSF} approach provides a reduction of 
38\%, 36.2\% and 28.7\% as compared to the considered temporal methods (persistence forecasting, \ac{AR} of order 3, and \ac{WT}-\ac{ANN} models) and a reduction of 24\% and 20\% as compared to the considered spatio-temporal methods (ANN-based ST and \ac{LS}-based ST), respectively. 
\begin{table}[tb]
\centering
\caption{Statistical Error Measure Comparison of Different Methods}
\scriptsize
\begin{tabular}{lcccccc}
\toprule
\textbf{Forecasting approach} & \textbf{MAE (m/s) } & \textbf{RMSE (m/s)} & \textbf{NRMSE (\%)}\\
\midrule
\textbf{Persistence Forecasting} &  2.1441 & 2.8334 & 16.86\\
\textbf{AR of order 1} & 2.0742 & 2.7629 & 16.44\\
\textbf{AR of order 3} & 2.0696 &  2.7560 & 16.40 \\
\textbf{WT-ANN} & 1.8200 & 2.4671 & 14.68\\
\textbf{ANN-based ST} & 1.7981 & 2.2997 & 13.69\\
\textbf{LS-based ST} & 1.7234 & 2.1983 & 13.08\\
\textbf{CST-WSF of order 3} &1.5665 & 2.0595 & 12.25\\
\textbf{Nonuniform CST-WSF} &1.3442 & 1.7586 & 10.46\\
\bottomrule
\end{tabular}
\label{error measures}
\end{table}

\begin{figure}[ht!]
\centering
\subfigure[]{
   \includegraphics[width = 0.825\columnwidth] {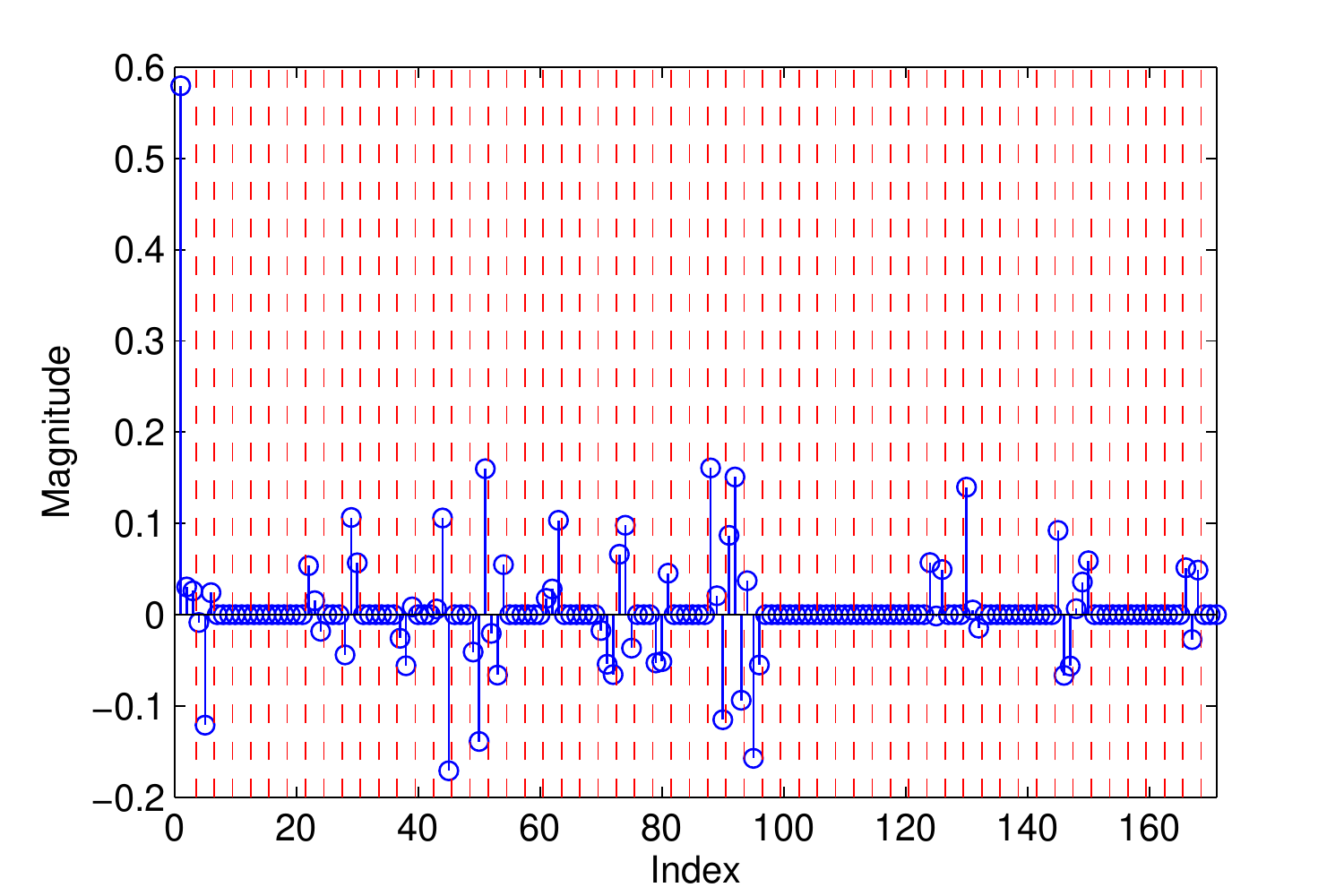}
   \label{}
 }
 \subfigure[]{
   \includegraphics[width = 0.825\columnwidth] {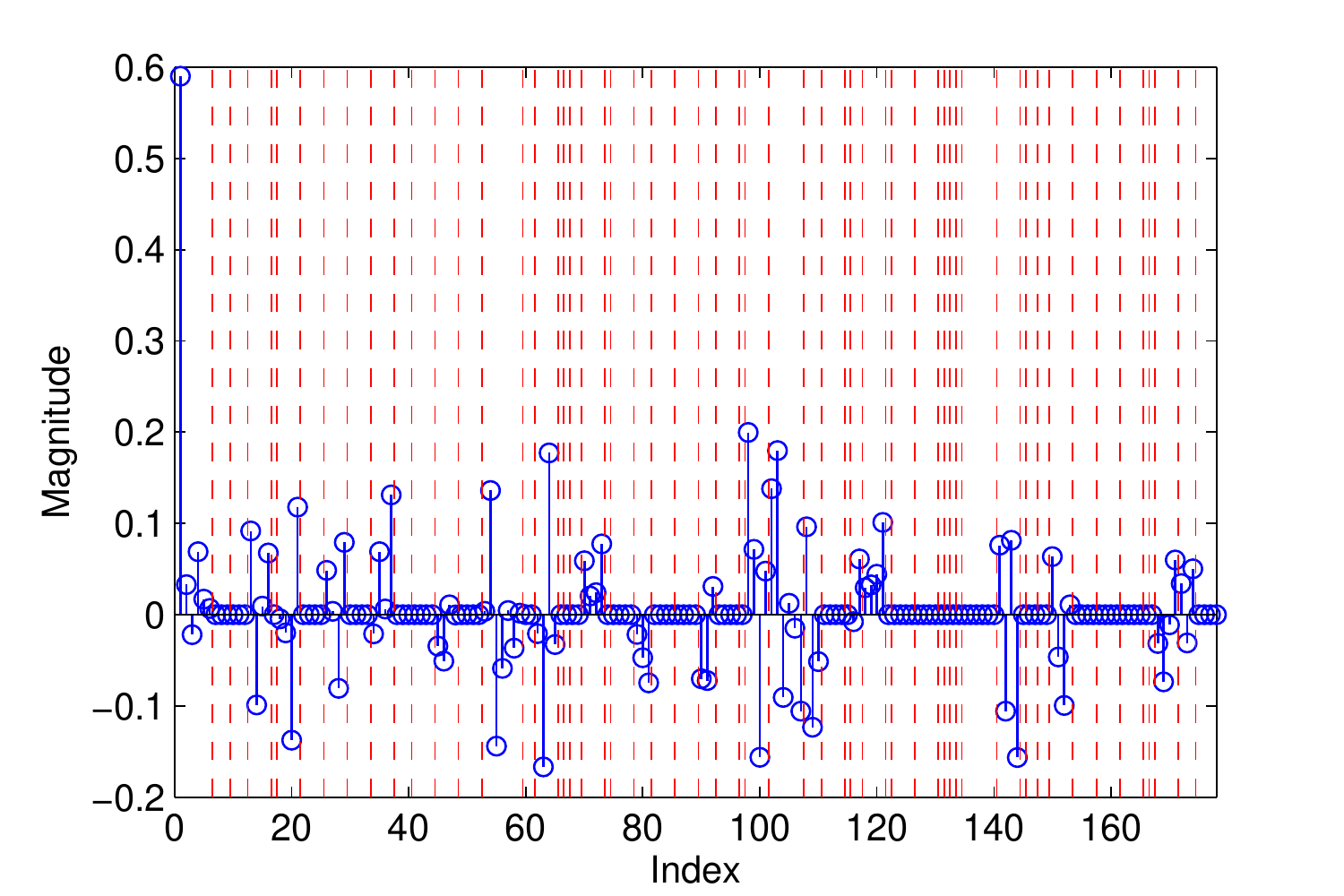}
   \label{}
 }
\caption{Block-sparse coefficient vector. The red dashed lines specify 57 vector-blocks of the coefficient vector. (a) Uniform CST-WSF of order 3. (b) Nonuniform CST-WSF with different orders (vector-block lengths).}
\label{fig:X_vectors}
\end{figure}

Fig. \ref{fig:X_vectors} illustrates the corresponding block-sparse coefficient vectors for the uniform and nonuniform \ac{CST-WSF} methods at the training stage.
As can be seen, only a few of the blocks in uniform and nonuniform \ac{CST-WSF} are non-zero, resulting in a block-sparse $\vc{x}$. This block-sparse structure  appears in all of the other calculated coefficient vectors (with different block-sparsity pattern) as we move over prediction horizon time and further confirms our motivation behind exploiting the intrinsic low-dimensional models in spatio-temporal wind speed forecasting. 
It is worth noting that the proposed \ac{CST-WSF} methods have a much shorter computational time as compared to other \ac{ANN}-based methods and the average computational time for other proposed short-term forecasting methods in the literature~\cite{xieshort,tascikaraoglu2014demand}.       
For instance, the total simulation time of nonuniform \ac{CST-WSF} approach is almost the half of the time required for the predictions with \ac{WT}-\ac{ANN} model and approximately 15\% smaller than that of \ac{LS} \ac{M-AR} spatio-temporal model in this study. 


\section{Conclusion}
\label{sec:conc}

We proposed two spatio-temporal wind speed forecasting methods, called uniform and nonuniform \ac{CST-WSF}.
The methods are inspired by \ac{CS} and structured-sparse recovery algorithms, where we claim that there usually exists an intrinsic low-dimensional structure governing a large collection of stations.
%
%
The results of a case study show that the proposed approaches considerably improves the short-term forecasts compared to a set of widely-used benchmark models.    

As future directions, we plan to apply the proposed \ac{CST-WSF} to a much larger set of stations. Incorporating other variables (such as temperature, pressure, etc.) in the wind speed forecasting is another research path. Yet another direction is to investigate using sparsity-based ideas in probabilistic forecasting methods.
Such information about the forecast errors are useful for \ac{TSOs}, \ac{IPPs}, etc., in evaluating the economic and technical risks due to uncertainty.


\bibliographystyle{IEEEtran}
\bibliography{reference_ACC_2014_Forecast.bib}

\begin{thebibliography}{10}
\providecommand{\url}[1]{#1}
\csname url@rmstyle\endcsname
\providecommand{\newblock}{\relax}
\providecommand{\bibinfo}[2]{#2}
\providecommand\BIBentrySTDinterwordspacing{\spaceskip=0pt\relax}
\providecommand\BIBentryALTinterwordstretchfactor{4}
\providecommand\BIBentryALTinterwordspacing{\spaceskip=\fontdimen2\font plus
\BIBentryALTinterwordstretchfactor\fontdimen3\font minus
  \fontdimen4\font\relax}
\providecommand\BIBforeignlanguage[2]{{%
\expandafter\ifx\csname l@#1\endcsname\relax
\typeout{** WARNING: IEEEtran.bst: No hyphenation pattern has been}%
\typeout{** loaded for the language `#1'. Using the pattern for}%
\typeout{** the default language instead.}%
\else
\language=\csname l@#1\endcsname
\fi
#2}}

\bibitem{iea2013technology}
\BIBentryALTinterwordspacing
{International Energy Agency (IEA)}, ``Technology roadmap: Wind energy,'' 2013.
  [Online]. Available:
  \url{http://www.iea.org/publications/freepublications/publication/Wind_2013_Roadmap.pdf}
\BIBentrySTDinterwordspacing

\bibitem{CA_renewable_portfolio}
\BIBentryALTinterwordspacing
{California Energy Commission}, ``California renewable energy overview and
  programs,'' 2013. [Online]. Available:
  \url{http://www.energy.ca.gov/renewables/index.html}
\BIBentrySTDinterwordspacing

\bibitem{smith2007utility}
J.~C. Smith, M.~R. Milligan, E.~A. DeMeo, and B.~Parsons, ``Utility wind
  integration and operating impact state of the art,'' \emph{IEEE Transactions
  on Power Systems}, vol.~22, no.~3, pp. 900--908, 2007.

\bibitem{bitar2011role}
E.~Bitar, R.~Rajagopal, P.~Khargonekar, and K.~Poolla, ``The role of co-located
  storage for wind power producers in conventional electricity markets,''
  \emph{Proceedings of the $2011$ American Control Conference}, pp. 3886--3891,
  2011.

\bibitem{callaway2009tapping}
D.~S. Callaway, ``Tapping the energy storage potential in electric loads to
  deliver load following and regulation, with application to wind energy,''
  \emph{Energy Conversion and Management}, vol.~50, no.~5, pp. 1389--1400,
  2009.

\bibitem{tascikaraoglu2011power}
A.~Tascikaraoglu, M.~Uzunoglu, B.~Vural, and O.~Erdinc, ``Power quality
  assessment of wind turbines and comparison with conventional legal
  regulations: A case study in turkey,'' \emph{Applied Energy}, vol.~88, no.~5,
  pp. 1864--1872, 2011.

\bibitem{sanandaji2014fast}
B.~M. Sanandaji, H.~Hao, and K.~Poolla, ``Fast regulation service provision via
  aggregation of thermostatically controlled loads,'' \emph{Proceedings of the
  $47$th Hawaii International Conference on System Sciences}, pp. 2388--2397,
  2014.

\bibitem{qin2014modeling}
J.~Qin, Y.~Chow, J.~Yang, and R.~Rajagopal, ``Modeling and online control of
  generalized energy storage networks,'' \emph{Proceedings of $5$th
  International Conference on Future Energy Systems}, pp. 27--38, 2014.

\bibitem{wang2008security}
J.~Wang, M.~Shahidehpour, and Z.~Li, ``Security-constrained unit commitment
  with volatile wind power generation,'' \emph{IEEE Transactions on Power
  Systems}, vol.~23, no.~3, pp. 1319--1327, 2008.

\bibitem{tascikaraoglu2014adaptive}
A.~Tascikaraoglu, O.~Erdinc, M.~Uzunoglu, and A.~Karakas, ``An adaptive load
  dispatching and forecasting strategy for a virtual power plant including
  renewable energy conversion units,'' \emph{Applied Energy}, vol. 119, pp.
  445--453, 2014.

\bibitem{zhu2012short}
X.~Zhu and M.~G. Genton, ``Short-term wind speed forecasting for power system
  operations,'' \emph{International Statistical Review}, vol.~80, no.~1, pp.
  2--23, 2012.

\bibitem{tascikaraoglu2014review}
A.~Tascikaraoglu and M.~Uzunoglu, ``A review of combined approaches for
  prediction of short-term wind speed and power,'' \emph{Renewable and
  Sustainable Energy Reviews}, vol.~34, pp. 243--254, 2014.

\bibitem{gneiting2006calibrated}
T.~Gneiting, K.~Larson, K.~Westrick, M.~G. Genton, and E.~Aldrich, ``Calibrated
  probabilistic forecasting at the stateline wind energy center: {T}he
  regime-switching space--time method,'' \emph{Journal of the American
  Statistical Association}, vol. 101, no. 475, pp. 968--979, 2006.

\bibitem{hering2010powering}
A.~S. Hering and M.~G. Genton, ``Powering up with space-time wind
  forecasting,'' \emph{Journal of the American Statistical Association}, vol.
  105, no. 489, pp. 92--104, 2010.

\bibitem{xieshort}
L.~Xie, Y.~Gu, X.~Zhu, and M.~G. Genton, ``Short-term spatio-temporal wind
  power forecast in robust look-ahead power system dispatch,'' \emph{IEEE
  Transactions on Smart Grid}, vol.~5, no.~1, pp. 511--520, 2014.

\bibitem{dowell2013short}
J.~Dowell, S.~Weiss, D.~Hill, and D.~Infield, ``Short-term spatio-temporal
  prediction of wind speed and direction,'' \emph{Wind Energy}, 2013.

\bibitem{he2014spatio}
M.~He, L.~Yang, J.~Zhang, and V.~Vittal, ``A spatio-temporal analysis approach
  for short-term forecast of wind farm generation,'' \emph{IEEE Transactions on
  Power Systems}, vol.~29, no.~4, pp. 1611--1622, 2014.

\bibitem{tastu2011spatio}
J.~Tastu, P.~Pinson, E.~Kotwa, H.~Madsen, and H.~A. Nielsen, ``Spatio-temporal
  analysis and modeling of short-term wind power forecast errors,'' \emph{Wind
  Energy}, vol.~14, no.~1, pp. 43--60, 2011.

\bibitem{tastu2014probabilistic}
J.~Tastu, P.~Pinson, P.-J. Trombe, and H.~Madsen, ``Probabilistic forecasts of
  wind power generation accounting for geographically dispersed information,''
  \emph{IEEE Transactions on Smart Grid}, vol.~5, no.~1, pp. 480--489, 2014.

\bibitem{wytock2013largescale}
M.~Wytock and Z.~Kolter, ``Large-scale probabilistic forecasting in energy
  systems using sparse gaussian conditional random fields,'' \emph{Proceedings
  of the $52$nd IEEE Conference on Decision and Control}, pp. 1019--1024, 2013.

\bibitem{zhang2014review}
Y.~Zhang, J.~Wang, and X.~Wang, ``Review on probabilistic forecasting of wind
  power generation,'' \emph{Renewable and Sustainable Energy Reviews}, vol.~32,
  pp. 255--270, 2014.

\bibitem{candes2006robust}
E.~J. Cand{\`e}s, J.~Romberg, and T.~Tao, ``{Robust uncertainty principles:
  Exact signal reconstruction from highly incomplete frequency information},''
  \emph{IEEE Transactions on Information Theory}, vol.~52, no.~2, pp. 489--509,
  2006.

\bibitem{tropp2004greed}
J.~Tropp, ``Greed is good: Algorithmic results for sparse approximation,''
  \emph{IEEE Transactions on Information Theory}, vol.~50, no.~10, pp.
  2231--2242, 2004.

\bibitem{donoho2001uncertainty}
D.~Donoho and X.~Huo, ``Uncertainty principles and ideal atomic
  decomposition,'' \emph{IEEE Transactions on Information Theory}, vol.~47,
  no.~7, pp. 2845--2862, 2001.

\bibitem{baraniuk2010mbcs}
R.~G. Baraniuk, V.~Cevher, M.~F. Duarte, and C.~Hegde, ``Model-based
  compressive sensing,'' \emph{IEEE Transactions on Information Theory},
  vol.~56, no.~4, pp. 1982--2001, 2010.

\bibitem{eldar2010block}
Y.~C. Eldar, P.~Kuppinger, and H.~B\"{o}lcskei, ``Block-sparse signals:
  uncertainty relations and efficient recovery,'' \emph{IEEE Transactions on
  Signal Processing}, vol.~58, no.~6, pp. 3042--3054, 2010.

\bibitem{sanandaji2012review}
B.~M. Sanandaji, T.~L. Vincent, and M.~B. Wakin, ``A review on sufficient
  conditions for structure identification of interconnected systems,''
  \emph{Proceedings of the $16$th IFAC Symposium on System Identification}, pp.
  1623--1628, 2012.

\bibitem{sanandaji2012thesis}
B.~M. Sanandaji, ``Compressive system identification ({CSI}): {T}heory and
  applications of exploiting sparsity in the analysis of high-dimensional
  dynamical systems,'' Ph.D. dissertation, Colorado School of Mines, 2012.

\bibitem{Iowa}
\BIBentryALTinterwordspacing
{Iowa Environmental Mesonet}, ``{ASOS} historical data,'' 2014. [Online].
  Available: \url{http://mesonet.agron.iastate.edu/ASOS/}
\BIBentrySTDinterwordspacing

\bibitem{tascikaraoglu2012assessment}
A.~Tascikaraoglu, M.~Uzunoglu, and B.~Vural, ``The assessment of the
  contribution of short-term wind power predictions to the efficiency of
  stand-alone hybrid systems,'' \emph{Applied Energy}, vol.~94, pp. 156--165,
  2012.

\bibitem{bilgili2007application}
M.~Bilgili, B.~Sahin, and A.~Yasar, ``Application of artificial neural networks
  for the wind speed prediction of target station using reference stations
  data,'' \emph{Renewable Energy}, vol.~32, no.~14, pp. 2350--2360, 2007.

\bibitem{tascikaraoglu2014demand}
A.~Tascikaraoglu, A.~Boynuegri, and M.~Uzunoglu, ``A demand side management
  strategy based on forecasting of residential renewable sources: A smart home
  system in turkey,'' \emph{Energy and Buildings}, vol.~80, pp. 309--320, 2014.

\end{thebibliography}

\end{document}